# Compact transient-grating self-referenced spectral interferometry for sub-nanojoule femtosecond pulses characterization

Xiong Shen,[1,2] Peng Wang,[1,2] Jun Liu,[1, 3, a)] and Ruxin Li[1,3]

[1)]State Key Laboratory of High Field Laser Physics, Shanghai Institute of Optics and Fine Mechanics, Chinese Academy of Sciences, Shanghai 201800, China

[2)]University of Chinese Academy of Sciences, Beijing 100049, China

[3)]IFSA Collaborative Innovation Center, Shanghai Jiao Tong University, Shanghai 200240, China

[a)] Corresponding email: jliu@siom.ac.cn

**Abstract:** The self-referenced spectral interferometry (SRSI) technique, which is usually used for microjoule-level femtosecond pulses characterization, is improved to characterize weak femtosecond pulses with nanojoule based on the transient-grating effect. Both femtosecond pulses from an amplifier with 3 nJ per pulse at 1 kHz repetition rates and femtosecond pulses from an oscillator with less than 0.5 nJ per pulse at 84 MHz repetition rates are successfully characterized. Furthermore, through a special design, the optical setup of the device is even smaller than a palm which will makes it simple and convenient during the application. These improvements extend the application of SRSI technique to the characterization of femtosecond pulses in a broad range. Not only pulses from an amplifier but also pulses from an oscillator or weak pulses used in ultrafast spectroscopy can be monitored with this SRSI method right now.

Femtosecond laser pulses have been widely used as a powerful tool in lots of scientific research fields in physics, material, chemistry, and biology. Methods for the characterization of femtosecond pulses improved as the improvement of femtosecond laser technology in the past decades. The step of pursuing a fast, accurate method with an extremely simple setup for femtosecond pulses characterization never stop. So far, the most widely used techniques for femtosecond pulses characterization are frequency-resolved optical gating (FROG) [1], spectral phase interferometry for direct electric-field reconstruction (SPIDER) [2] and Self-referenced spectral interferometry technique (SRSI) [3]. FROG, which was appeared in 1993, usually needs a relatively long time for iterative calculation to retrieve the spectral phase and the temporal profile. SPIDER, which was shown in 1998, has a relatively complicated optical setup due to needing a chirped pulse. SRSI was shown in 2010 with several advantages in comparison to FROG and SPIDER it has turned out to be a linear, analytical, sensitive, accurate and fast technique [3].In brief, the principle of SRSI can be expressed as following: A reference pulse was generated from the unknown input pulse itself by using a frequency-conserving nonlinear optical effect firstly, then the self-created reference pulse collinearly mixed with the unknown input pulse that separated by a time delay τ and results in a spectral interferogram in a spectrometer, the spectral amplitude and phase of the unknown input pulse can then be retrieved from the spectral interferogram by Fourier-transform spectral interferometry (FTSI) [4]. SRSI hugely simplifies the femtosecond pulses measurement setup and the retrieving algorithm [5].

Pulses up to 50 dB dynamic range can be characterized by using SRSI even with a common fiber spectrometer. As a result, this method has been used in many applications where a real time monitor, a single-shot characterization, or feedback control and optimization of an input pulse are needed [6-8].

In SRSI, in order to generate an appropriate reference pulse, a frequency-conserving third order nonlinear process is needed. Up to now, three frequency-conserving nonlinear processes have been used in the SRSI technique, they are cross-polarized wave generation (XPW) [3], self-diffraction (SD) [9] and transient grating (TG) [10]. Among them, the XPW process is the first used and widely explored process for SRSI method [11-14]. The use of third-order nonlinear optical processes, which needs relatively high incident pulse energy to generate reference pulses, precludes the SRSI technique for weak femtosecond pulse characterization. For example, with the SRSI technique, a pulse with typically about 1 µJ pulse energy is required for the characterization of a 50 fs pulse based on the XPW generation [5]. As second order nonlinear optical processes, such as second-harmonic generation or sum-frequency generation, can be used in the FROG and SPIDER technique, both of these techniques do better in weak femtosecond pulse characterization than SRSI.

As we know, weak femtosecond pulses generated from oscillators or noncollinear optical parametric amplifiers etc, with sub-nanojoule to several nanojoule energies, are widely used to explore ultrafast phenomena in biology and chemistry. Accurate and complete temporal characterization of such pulses is of vital important in these applications. To extend the application range of SRSI method to weak pulses characterization, the TG based SRSI technique is explored here. Through a special design, we succeed an extremely simple optical setup based on this method. Femtosecond pulses with 3 nJ at 1 kHz repetition rates and femtosecond pulses from an 84 MHz oscillator with less than 0.5 nJ energy per pulse are successfully characterized. The results drastically improve the application of SRSI in the characterization and monitoring of weak femtosecond pulses that used in ultrafast spectroscopy etc.

For a successful SRSI characterization, it is of vital important to self-create a reference pulse which has a shorter duration and a broader spectrum with nearly negligible phase variations compared with the unknown input pulse. The TG process is self-phase matched at all wavelengths, background free, and alignment free as the reference pulse is created in the same direction of the unknown input pulse to be characterized [10]. What is more, TG process roused an interest on weak pulse measurement owing to its relative higher energy sensitivity [15] in comparison to the SD and XPW processes. Previous work showed femtosecond pulses with less than 100 nJ pulse energy had been successfully characterized [16]. However, it is still far from sub-nanojoule level and can not be used for the characterization of femtosecond pulses from oscillators. Then, how can we improve the method to measure a two orders of energy lower femtosecond pulse? How can we simplify the optical setup?

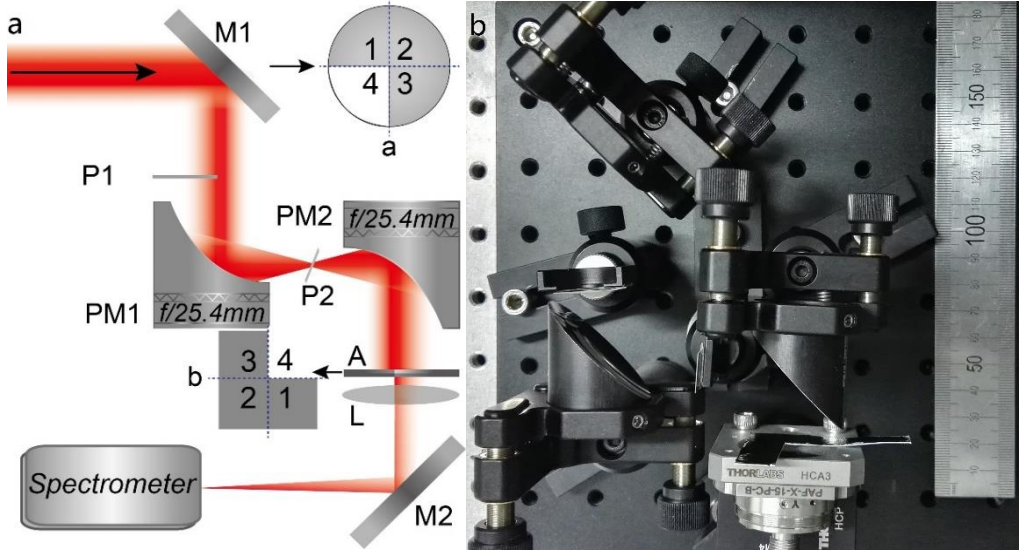

Fig. 1. Experimental setup. Fig. 1(a), M1, special 3/4 coated reflective mirror; P1, fused silica plate, 0.5 mm thick; PM1, PM2, 90° off-axis parabolic aluminum mirror, f=25.4 mm; P2, YAG plate, 0.15 mm thick; A, aperture; L, lens, f=200 mm; M2, reflective mirror; Inset a, the front side of M1; Inset b, the shape of A. Fig. 1(b), the picture of the optical setup of the device.

Three key points are applied to successfully characterize sub-nanojoule femtosecond pulses. The first is that TG process with higher energy sensitivity is used to generate the reference pulse. The second point is the nonlinear medium with relatively higher third-order nonlinearity is chosen for the TG process. The third and also the most important point is that a symmetric focusing and collimation pair of parabolic mirrors with a tight focal length are used to increase the focal intensity in the nonlinear medium as high as possible. The symmetric pair of parabolic mirrors also simplify the setup and make the setup easy to be aligned.

The schematic of the TG-SRSI experimental setup is shown in Fig. 1. The unknown input pulse is reflected by a special coated mirror M1 first. The front side of M1 is shown in inset a in Fig. 1 in which the 3/4 area of its front surface is high reflectivity coated and the rest 1/4 area is not coated. After the M1 mirror, the unknown input pulse beam can be regarded as four beams: $E_1(\omega)$, $E_2(\omega)$, $E_3(\omega)$ and $E_4(\omega)$ which reflected from the areas 1, 2, 3 and 4 of M1, respectively. Obviously, the first three beams can be looked as three identical incident beams for TG process, while $E_4(\omega)$ is dramatically attenuated due to the uncoated glass surface and used as the unknown input pulse to be calculated. An off-axis parabolic aluminum mirror PM1 with only 25.4 mm effective focal length is used to focus the four beams. A TG reference pulse $E_{ref}(\omega)$ is generated when $E_1(\omega)$, $E_2(\omega)$, $E_3(\omega)$ are focused in a nonlinear plate P2 simultaneously owning to transient-grating (TG) nonlinear process. P2 is a YAG plate with a thickness of 0.15 mm. Then the reference pulse $E_{ref}(\omega)$ can be expressed as:

$$E_{ref}(\omega) \propto E_1(\omega)E_2^*(\omega)E_3(\omega) = E_{in}^2(\omega)E_{in}^*(\omega). \tag{1}$$

In time domain, the intensity of the TG signal can be denoted as:

$$I_{ref}(\Omega,t) \propto \left|\iint d\omega' d\omega'' E_1(t,\omega')E_2^*(t,\omega'')E_3(t,\Omega-\omega''+\omega')\right|^2. \quad (2)$$

The time domain expression indicates a shorter reference pulse generation, and the reference pulse has a smoother and broader spectrum than the unknown input pulse as it takes into account all the frequency combinations that contribute to the signal at a given frequency. What is more, the cubic dependence of the third order TG nonlinear process can reduces the input second order spectral phase by a factor as large as 9, which makes the phase of the reference pulse smooth or even flat [17].

Beam $E_4(\omega)$ will be characterized as the unknown input pulse, let's denote it as $E_{unk}(\omega)$. $E_{unk}(\omega)$ is attenuated by the reflection of the 1/4 uncoated area of M1, when the incident angle on M1 of the p-polarized unknown input pulse is about 45 degree, about $(1/4 \times 0.6533)\% \approx 0.16\%$ energy of the unknown input pulse is kept in this beam after reflection of M1. A 0.5 mm fused silica plate P1 is inserted in the propagation path of $E_{unk}(\omega)$ to introduce a time delay $\tau$ between $E_{unk}(\omega)$ and $E_{ref}(\omega)$, it can also attenuate $E_{unk}(\omega)$ to a certain value when coated with different reflection ratio.

After P2, another off-axis parabolic aluminum mirror PM2 with the same 25.4 mm effective focal length is used to symmetrically collimate the beam. The two beam $E_{unk}(\omega)$ and $E_{ref}(\omega)$ are in the same direction automatically owing to the BOXCARS beam geometry in the TG process [10, 18]. Both the symmetric parabolic mirrors pair and the BOXCARS geometry make the whole setup easy to be aligned. The two beams $E_{unk}(\omega)$ and $E_{ref}(\omega)$ with the time delay $\tau$ are filtered out by an aperture A (inset b in Fig. 1) and guided to a spectrometer (HR 4000, Ocean Optics) through a lens L with 200 mm focal length and a high reflectivity mirror M2. The aperture A can be adjusted appropriately to eliminate the affecting of scattering light caused by the two dividing edges between the coated and uncoated areas in M1. An interference spectrum of $E_{unk}(\omega)$ and $E_{ref}(\omega)$ measured by the spectrometer can then ben expressed as:

$$D(\omega,\tau) = \left|E_{ref}(\omega) + E_{unk}(\omega)e^{i\omega\tau}\right|^2$$
$$= \left|E_{ref}(\omega)\right|^2 + \left|E_{unk}(\omega)\right|^2 + E_{ref}^*(\omega)E_{unk}(\omega)e^{i\omega\tau} + E_{ref}(\omega)E_{unk}^*(\omega)e^{-i\omega\tau} \quad (3)$$

where $S_0(\omega) = \left|E_{ref}(\omega)\right|^2 + \left|E_{unk}(\omega)\right|^2$ is the sum of the spectra between the reference pulse and the unknown input pulse, $f(\omega) = E_{ref}^*(\omega)E_{unk}(\omega)$ is the interference part of the two pulses.

The spectral amplitudes of the reference pulse and the unknown input pulse and the spectral phase difference between these two pulses can be retrieved by FTSI directly

$\left|\tilde{E}(\omega)\right| = \frac{1}{2}(\sqrt{\tilde{S}_0(\omega) + 2\left|\tilde{f}(\omega)\right|} - \sqrt{\tilde{S}_0(\omega) - 2\left|\tilde{f}(\omega)\right|})$ , $\left|\tilde{E}_{ref}(\omega)\right| = \frac{1}{2}(\sqrt{\tilde{S}_0(\omega) + 2\left|\tilde{f}(\omega)\right|} + \sqrt{\tilde{S}_0(\omega) - 2\left|\tilde{f}(\omega)\right|})$ , and

$\phi_{unk}(\omega) = \phi_{ref}(\omega) - \arg(\tilde{f}(\omega)) \approx -\arg(\tilde{f}(\omega))$ owing to $\phi_{ref}(\omega)$ is almost negligible.

The device is firstly applied to the characterization of a femtosecond pulse at 1 kHz repetition rates. A small portion of a 48 fs/4 mJ pulse centered at 800 nm, generated from a Ti: sapphire laser system (Legend Elite Series, Coherent) is guided to the TG-SRSI device. The diameter of the input beam is about 10 mm. A variable neutral attenuation plate is used to adjust the input pulse energy, the characterization results of pulses with energy of 5 nJ, 4 nJ and 3 nJ are chosen for comparison.

Figure 2 shows the results of the measurement. As the characterization results of the laser pulse at three different pulse energies are almost the same, only the obtained spectral phase and temporal profile from the 3 nJ incident pulse are shown in Fig. 2 (a) and Fig. 2 (b).

For SRSI, there are two simple and direct standards for a correct measurement, one is that the self-created reference pulse is broader than the unknown input pulse, the other is that the measured spectra match the retrieved spectra.

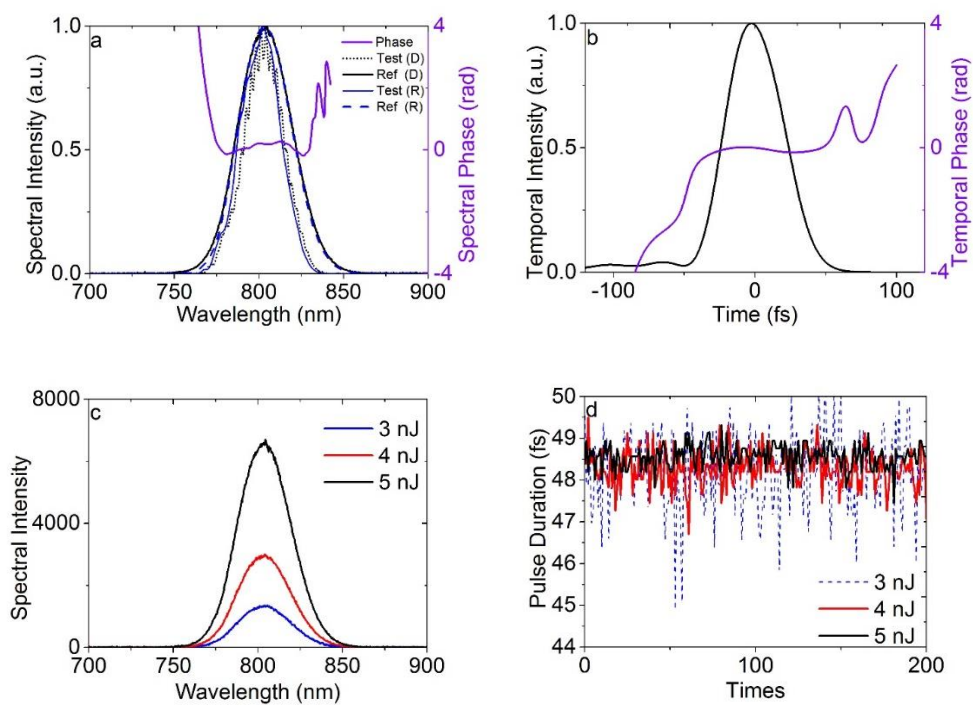

Fig. 2. The characterization results of femtosecond pulses with 5 nJ, 4 nJ and 3 nJ pulse energy at 1 kHz repetition rates. Fig. 2 (a), the retrieved and directly measured spectra of the unknown input pulse and TG reference when with input pulse energy of 3 nJ. The retrieved spectral phase is also shown as thick solid purple line. Fig. 2 (b), the retrieved temporal profile and phase of the unknown input pulse with input energy of 3 nJ. Fig. 2 (c), the spectra intensity of self-created TG reference pulses with input energy of 5 nJ, 4 nJ and 3 nJ, respectively. Fig. 2 (d), the pulse duration fluctuation in the characterization when the input pulse energy are 5 nJ, 4 nJ and 3 nJ, respectively.

In Fig. 2 (a), it can be seen that the spectrum of the self-created reference pulse (thick solid black line) is broader and smoother than the spectrum of the unknown input pulse (thin dotted black line) measured by spectrometer directly, which is of vital important in SRSI characterization. What is more, the retrieved spectra of the unknown input pulse (thin solid blue line) and the reference pulse (thick dashed blue line) are in coincidence well with the measured spectra of the unknown input pulse (thin dotted black line) and the reference pulse (thick solid black line), respectively. This self-checked result verifies the reliability of the device.

Figure 2 (b) shows the retrieved temporal profile and phase of the unknown input pulse with input energy of 3 nJ.

Figure 2 (c) shows the spectral intensity of the self-created TG reference pulse at 5 nJ, 4 nJ and 3 nJ different energies. The spectral intensity correspond to 5 nJ, 4 nJ and 3 nJ incident pulse energy are about 6750, 3000 and 1350, respectively. The intensity ratio of the self-created TG reference pulses is $6750: 3000: 1350 = 5: 2.2: 1 = R$, and the intensity ratio of the three unknown input pules is $5^3: 4^3: 3^3 = 4.6: 2.3: 1 = U$, $R \approx U$, which keeps the relationship between the self-created reference pulse and the unknown input pulse as $I_{ref} \propto I_{unk}^3$, this result also verifies the reliability of the device

Figure 2 (d) shows the pulse duration fluctuation of the characterization result with different energy of 5 nJ, 4 nJ and 3 nJ. Each curve is consisted of 200 times characterization results. It is easy to acquire such curves as SRSI can realize real-time characterization. It is clearly shown that the fluctuation range increase with input pulse energy decrease. This is because of that effect from scattering noise and the spectrometer's dark noise to the characterization increase with the input pulse energy decrease.

Pulses with sub 1 nJ energy per pulse at 84 MHz repetition rates from a laser oscillator (Mai Tai SP, Spectral Physics) are also characterized. The pulse from the oscillator is first chirp compensated with a pair of chirped mirrors and then beam expanded to about 10 mm from 2 mm with a concave aluminum coated reflective mirror and a convex aluminum coated reflective mirror. A variable neutral attenuation plate is used to adjust the input pulse energy to the device.

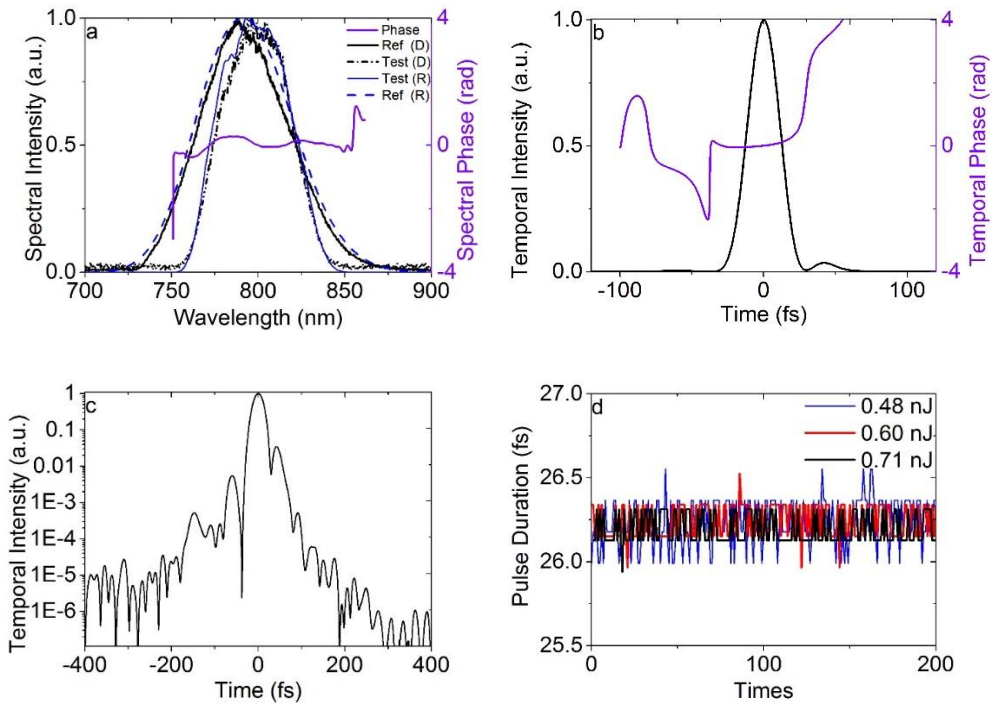

Fig. 3. The characterization results of femtosecond pulses with 0.48 nJ, 0.60 nJ and 0.71 nJ single pulse energy at 84 MHz repetition rates. Fig. 3 (a), the retrieved and directly measured spectra of the unknown input pulse and TG reference pulse when with input pulse energy of 0.48 nJ. The retrieved spectral phase is also shown as the thick solid purple line. Fig. 3 (b), the retrieved temporal profile and phase of the unknown input pulse with input pulse energy of 0.48 nJ. Fig. 3 (c), the retrieved temporal intensity profile on a logarithmic scale. Fig. 3 (d), the pulse duration fluctuation curves in the characterization when the input pulse energy are 0.48 nJ, 0.60 nJ and 0.71 nJ, respectively.

Figure 3 shows the characterization results of pulses with 0.48 nJ, 0.60 nJ and 0.71 nJ single pulse energy. The comparison of the retrieved spectra and the spectra measured by spectrometer directly are shown in Fig. 3 (a). The measured spectra match the retrieved spectra. The self-created reference pulse (thick solid black line) is broader and smoother than the spectrum of the input unknown input pulse (thin dash dotted black line). The pulse is well chirp compensated as the retrieved phase has little variation. Figure 3 (b) shows the retrieved temporal profile and phase of the unknown input pulse with input pulse energy of 0.48 nJ. Figure 3 (c) shows the retrieved temporal intensity profile on a logarithmic scale. The device shows a $10^5$ dynamic range on a $\pm$ 400 fs temporal range. The pulse duration is about 26.3 fs. Figure 3 (d) shows the pulse duration fluctuation of the characterization result with different energy of 0.48 nJ, 0.60 nJ and 0.71 nJ. Each curve is consisted of 200 times characterization results

In conclusion, we extend the SRSI technique to weak pulses characterization based on the transient-grating optical effect. By using a special design, the optical setup is easy to be aligned and even smaller than a palm. Less than 0.5 nJ pulses at 84 MHz and 3 nJ pulses at 1 kHz are successfully characterized with our extremely compact setup. In comparison to previous XPW and SD based SRSI, our novel TG based SRSI device increases the energy sensitivity from microjoule to sub-nanojoule by about three orders of magnitude. The results show that the novel TG-SRSI device can be used in the characterization and monitoring of the temporal profile of pulses from oscillators. It can also be used to characterize weak femtosecond pulses that used in ultrafast spectroscopy etc.

**Acknowledgements:**

This work is supported by the National Natural Science Foundation of China (NSFC) (grants 11274327, 61521093 and 61527821) and the Chinese Academy of Sciences (grant YZ201538).